\documentclass[10pt,conference]{IEEEtran}
\pdfoutput=1
\usepackage{cite}
\usepackage[pdftex]{graphicx}
\usepackage[cmex10]{amsmath}
\usepackage{array}

\usepackage{fixltx2e}

\usepackage{url}

\usepackage{color}
\definecolor{Red}{rgb}{1,0,0}

\usepackage{amsxtra}
\usepackage{amssymb}

\newcommand{\dT}{\ensuremath{\Delta T}}
\newcommand{\ket}[1]{| #1 \rangle}
\newcommand{\bra}[1]{\langle #1 |}

\newcommand{\braket}[2]{\langle #1 | #2 \rangle}
\newcommand{\magsqr}[1]{\left| #1 \right|^{2}}
\newcommand{\qop}[1]{\widehat{#1}}
\newcommand{\overlap}{s}
\newcommand{\pie}{c_p}
\newcommand{\die}{c_d}

\newcommand{\dieapproxa}{{}\tilde{c}_d^{(1)}}
\newcommand{\dieapproxb}{\tilde{c}_d^{(2)}}
\newcommand{\E}{E}       
\newcommand{\Nph}{E}   
\newcommand{\lo}{\text{lo}}
\newcommand{\codeset}{S_{\text{code}}}
\newcommand{\Pe}{P_e}
\newcommand{\Pev}{P_\text{even}}
\newcommand{\Podd}{P_\text{odd}}
\newcommand{\defn}{\equiv}
\newcommand{\beq}{\begin{equation}}
\newcommand{\eeq}{\end{equation}}
\newcommand{\eeql}[1]{\label{#1}\end{equation}}
\newcommand{\beqa}{\begin{IEEEeqnarray}{rCl}}
\newcommand{\eeqa}{\end{IEEEeqnarray}{rCl}}

\begin{document}
\title{On Approaching the Ultimate Limits of\\Photon-Efficient and Bandwidth-Efficient\\Optical Communication}

\author{\IEEEauthorblockN{Sam Dolinar, Kevin M. Birnbaum, Baris I. Erkmen, Bruce Moision}
\IEEEauthorblockA{
Jet Propulsion Laboratory, California Institute of Technology, Pasadena, CA, USA\\
email:\{sam.dolinar, kevin.m.birnbaum, baris.i.erkmen, bruce.moision\}@jpl.nasa.gov
}
}
\maketitle

\begin{abstract}
It is well known that ideal free-space optical communication at the quantum limit can have unbounded photon information efficiency (PIE), measured in bits per photon. High~PIE comes at a price of low dimensional information efficiency~(DIE), measured in bits per spatio-temporal-polarization mode.  If only temporal modes are used, then DIE translates directly to bandwidth efficiency.  
In this paper, the DIE vs. PIE tradeoffs for known modulations and receiver structures are compared to the ultimate quantum limit, and analytic approximations are found in the limit of high PIE.  This analysis shows that known structures fall short of the maximum attainable DIE by a factor that increases linearly with PIE for high PIE.

The capacity of the Dolinar receiver is derived for binary coherent-state modulations and computed for the case of on-off keying (OOK).
The DIE vs. PIE tradeoff for this case is improved only slightly compared to OOK with photon counting.
An adaptive rule is derived for an additive local oscillator that maximizes the mutual information between a receiver and a  
transmitter that selects from a set of coherent states.  For binary phase-shift keying (BPSK), this is shown to be equivalent to the operation of the Dolinar receiver.

The Dolinar receiver is extended to make {adaptive measurements} on a {coded sequence} of coherent state symbols. Information from previous measurements is used to adjust the {\em a~priori} probabilities of the next symbols. 
The adaptive Dolinar receiver does not improve the DIE vs. PIE tradeoff compared to independent transmission and Dolinar  
reception of each symbol.

\end{abstract}

\section{Introduction}
\begin{figure}[h]
\centering
\includegraphics[width=3.5in]{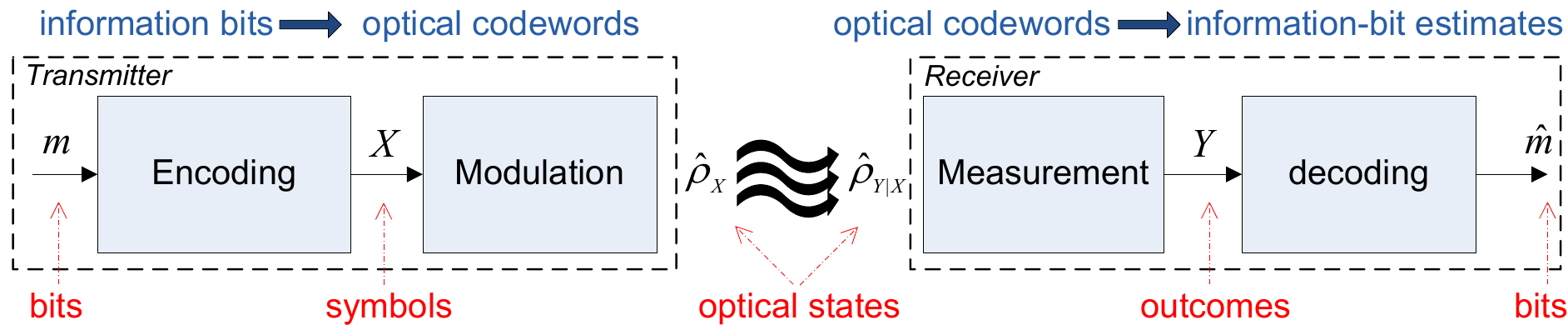}
\caption{Block diagram of an optical communication system. The transmitter performs encoding and modulation to map information bits into optical field states. The propagation maps these transmitted states into received states. The receiver performs a measurement on the received states and the outcomes are decoded to estimate the transmitted message.}
\label{OptCommBlockDiagram}
\end{figure}

	An optical communication system can be represented by the block diagram shown in Figure~\ref{OptCommBlockDiagram}.  At the transmitter structured redundancy is introduced into message bits (encoding), such that the receiver can correct errors introduced during communication. The encoded bits are then mapped into states of optical fields (modulation). These optical states are typically {\em coherent states}, which are produced by lasers; more generally, they could be thermal states or other more exotic quantum mechanical states. The optical fields convey the information through the communication medium (the \emph{channel}), and map to (a possibly different set of) states at the receiver. 
	For example, in ideal free-space communication, a coherent state is simply attenuated via propagation but otherwise unaltered, whereas in a turbulent channel it maps to a mixed state that is no longer a coherent state. The receiver performs a measurement on the received optical fields (demodulation or detection) and the measurement outcomes are then used to generate an estimate of the message that was transmitted (decoding). 
	
	In the classical formulation of information theory, the maximum rate of reliable communication is determined by maximizing Shannon's well-known mutual information metric~\cite{Shannon:CommThy,CoverThomas:InfThy}. In performing this maximization the channel is represented as a probability map, specifying the probability of a particular measurement outcome given the transmitted value. Thus, the mapping from bits to optical fields at the transmitter and the reverse mapping via the measurement at the receiver are part of the channel. Because light fields are fundamentally quantum mechanical, the probabilistic mapping defining the channel is determined by the \emph{choice of measurement}. For example, an ideal coherent-state measured with a photon-counting detector yields Poisson statistics with mean proportional to the incident power, whereas the same state measured with a homodyne receiver yields Gaussian statistics with mean equal to the field amplitude and variance~$1/4$~\cite{Shapiro:QuantumOptCommReview,NielsenChuang}. Hence, to find the highest rate of communication for a given modulation scheme, one must perform the daunting task of maximizing the Shannon information over \emph{all} possible measurements at the receiver. In the quantum-mechanical formulation of information theory, the highest rate of reliable communication is given by the 
\emph{Holevo information metric}~\cite{Holevo:Capacity,HausladenEtAl:Capacity,Schumacher:Capacity,NielsenChuang}, which implicitly maximizes the Shannon mutual information over all measurement schemes.
	
	The maximum rate of transmitting information reliably over a communication link is its \emph{capacity}, $C$, which is commonly measured as a channel throughput in units of bits per channel use.  It is also common to characterize the link's capability in terms of its \emph{efficiency} in utilizing link resources~\cite{Verdu:PhotonCost}.  The \emph{photon information efficiency} (\emph{PIE}), denoted by $\pie$, measures the efficiency of information transfer per unit \emph{energy}; it is given by:
	\begin{equation}
	\pie \equiv \frac{C}{E} \, \qquad \text{(bits per photon)},
	\end{equation}
where $E$ is an average photon number constraint imposed on the signal at the transmitter or receiver.   The \emph{dimensional information efficiency} (\emph{DIE}), denoted by $\die$, measures the efficiency of information transfer per unit \emph{dimension} of the optical signal; it is given by:
	\begin{equation}
	\die \equiv  \frac{C}{D} \, \qquad \text{(bits per dimension)},
	\end{equation}
where $D$ is the number of dimensions utilized per use of the channel. These dimensions can be any degree of freedom afforded to an optical communication link, namely temporal, spatial or polarization. 
In these expressions, the numbers of photons~$E$ and dimensions~$D$ are normalized per use of the channel, as is the capacity~$C$.
	
\begin{figure}[h]
\centering
\includegraphics[width=3.5in]{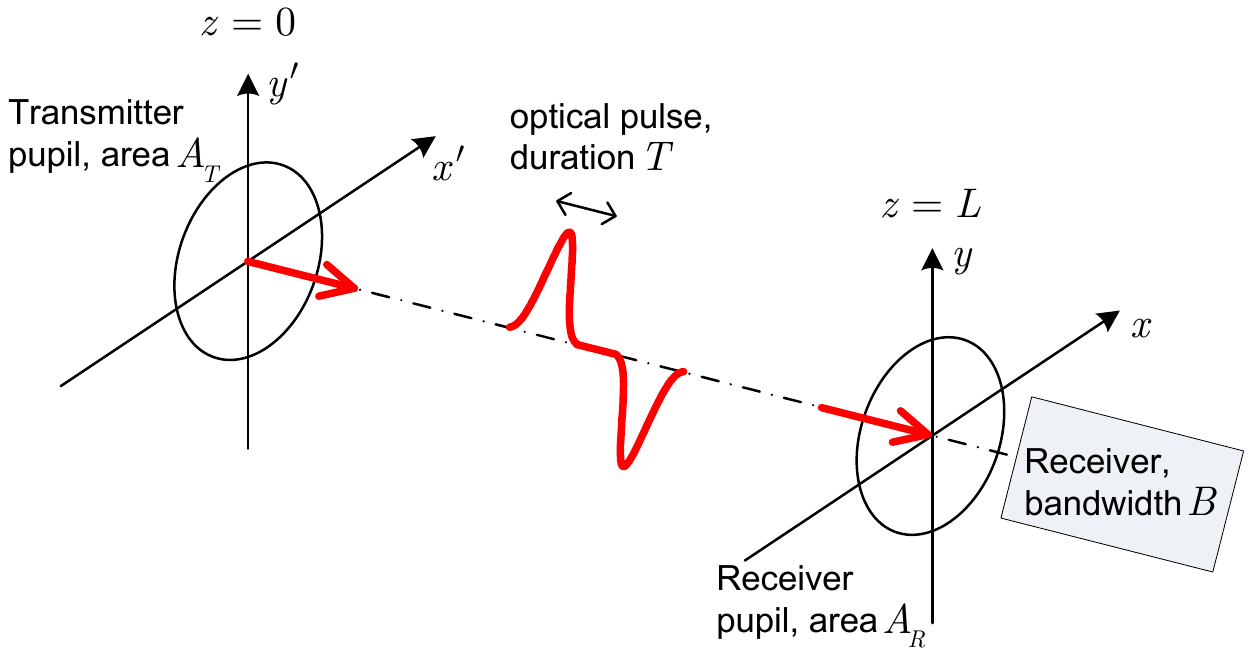}
\caption{The paraxial optical communication geometry, with time-bandwidth product $BT$ and Fresnel number product $A_{T} A_{R} /(\lambda L)^{2}$, where $\lambda$ is the center wavelength.}
\label{ParaxialGeometry}
\end{figure}

	As an example, consider a free-space (i.e., vacuum) optical communication link over a distance~$L$, utilizing transmitter and receiver apertures having unobscured areas $A_{T}$ and $A_{R}$, respectively (Figure~\ref{ParaxialGeometry}). 
	Let $\lambda$ denote the center wavelength of the optical fields, $T$ the duration of each optical pulse transmitted over the channel, and $B$ the total bandwidth occupied by the end-to-end system. The maximum number of independent temporal dimensions that could be utilized by this link for every use of the channel is approximately $BT$, and the maximum number of independent spatial dimensions is approximately $A_{T} A_{R}/(\lambda L)^2$~\cite{ShapiroGuhaErkmen:OptComm, Shapiro:QuantumOptCommReview}. 
If two independent polarizations are also utilized, then the maximum number of independent dimensions is approximately 
$D = 2 B T A_{T} A_{R} / (\lambda L)^2$, which can be thought of as the number of vector-valued spatio-temporal basis functions that are required to represent an arbitrary function in the (temporal, spatial, polarization) subspace determined by  the communication link parameters.

\section{Fundamental Capacity Efficiency Tradeoffs}
Practical laser systems typically generate coherent states using a convenient modulation such as pulse-position modulation (PPM), on-off keying (OOK), binary phase-shift keying (BPSK), or higher-order phase and/or amplitude modulations such as quadrature phase-shift keying (QPSK) and quadrature amplitude modulation (QAM).  It is well known that photon efficiency is unbounded in the quantum noise limit, and that this is achievable with a conventional system using PPM and photon counting.  But high photon efficiency comes at a price of exponentially decreasing dimensional efficiency. 
The ultimate Holevo limit is better than the Shannon limits based on PPM and other known strategies, but it still imposes harsh limitations on the dimensional efficiency that can be achieved at high photon efficiency.

\begin{figure}[t]
	\centering
	\includegraphics[width=3.5in]{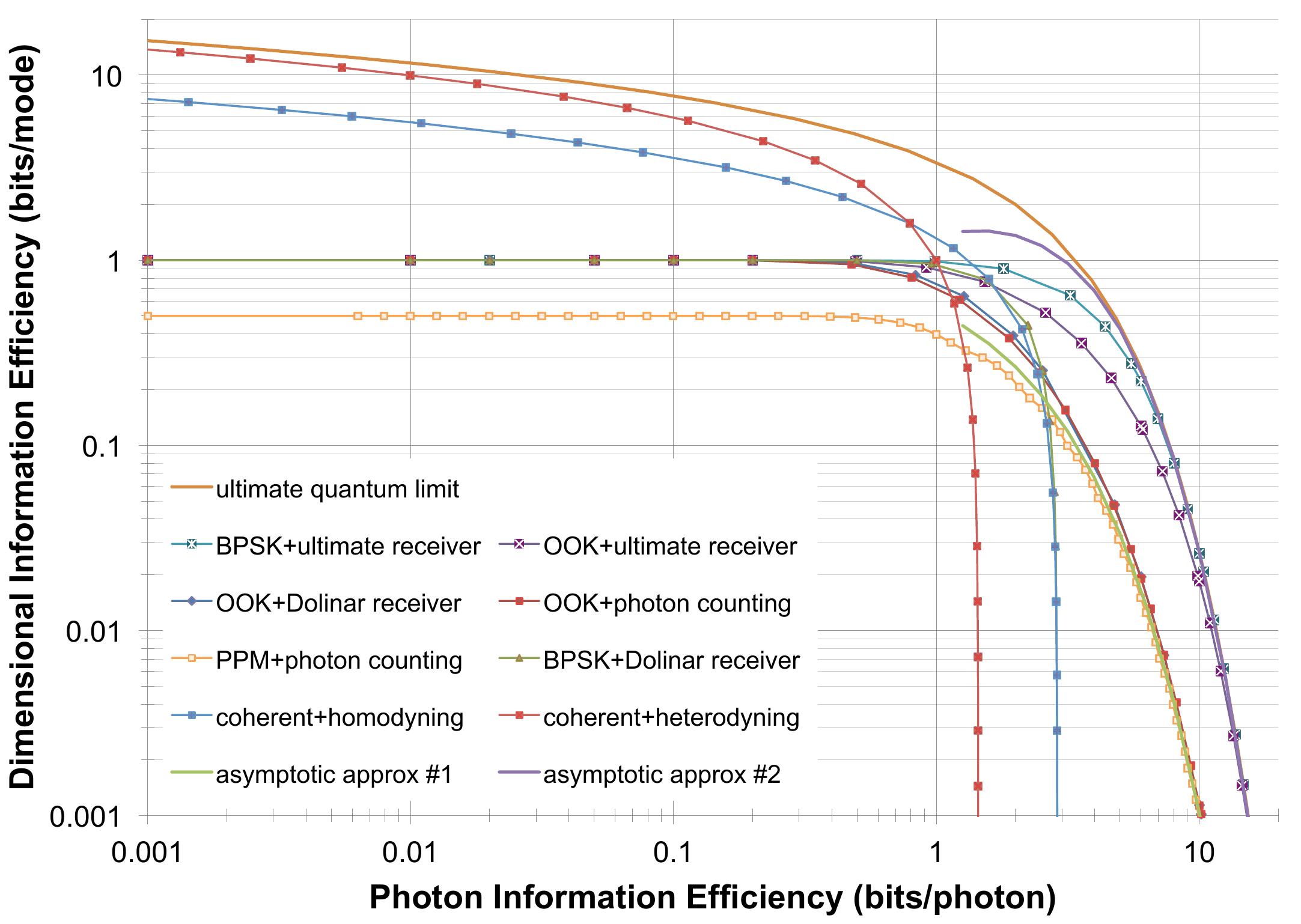}
	\caption{Theoretical limits on the best possible tradeoff between photon information efficiency (bits/photon) and dimensional information efficiency (bits/dimension or bits/mode).}
	\label{fig:DIEvsPIElimits}
\end{figure}

Our aim in this paper is to understand the gap between the Holevo limit and the Shannon limits of known systems, and to determine the characteristics of new systems that might bridge this gap.
Figure~\ref{fig:DIEvsPIElimits}  shows the fundamental DIE vs. PIE capacity tradeoffs for free-space optical communication under an average power constraint, under various assumptions about the optical modulation and detection.
All but four of the curves presented in Fig.~\ref{fig:DIEvsPIElimits} come from formulas stated 
in~\cite{ErkmenMoisionBirnbaum:IPNPR2009 , ErkmenMoisionBirnbaum:SPIE2010} and originally derived in references cited therein.  The curve labeled ``PPM + photon counting'' in Fig.~\ref{fig:DIEvsPIElimits} is based on an analytic approximation derived in Section~\ref{sec:CapAnalysis} of this paper, and it replaces a similar tradeoff curve computed numerically 
in~\cite{ErkmenMoisionBirnbaum:IPNPR2009 , ErkmenMoisionBirnbaum:SPIE2010}.  The curve labeled ``OOK + Dolinar receiver'' is new, and it comes from a capacity derivation for a general Dolinar receiver structure in Section~\ref{CapCohDR}.  Fig.~\ref{fig:DIEvsPIElimits} also shows two simple but accurate approximations (denoted as ``asymptotic approx~\#1'' and ``asymptotic approx~\#2'') that are derived in Section~\ref{sec:CapAnalysis} based on asymptotic analysis for high PIE of two of the tradeoff curves in Fig.~\ref{fig:DIEvsPIElimits}.

\subsection{The Ultimate Quantum Limit}
The ultimate quantum limit (outermost gold curve) is an immutable upper bound on the best possible tradeoff between photon efficiency and dimensional efficiency.  This limit can only be achieved with quantum-optimal modulation, detection, coding and decoding.    	For single-mode communication the dimensional efficiency (in bits/mode) on the vertical axis is the same as the spectral efficiency (in bits/sec/Hz).  The upper limit on the spectral efficiency of a multi-mode system scales in proportion to the number of independent non-temporal modes.  The ultimate quantum limit is achievable with optimal coherent-state 
modulation~\cite{GiovannettiGuhaEtAl:PureLossCapacity}. 

\subsection{Limits with Constrained Modulation and/or Detection}
 Fig.~\ref{fig:DIEvsPIElimits} also shows the (inferior) tradeoff limits obtained by restricting the modulation and/or detection to sub-optimal forms.  Heterodyne or homodyne receivers can be used with arbitrary coherent-state modulation, and such receivers teamed with high-order modulations can achieve high dimensional efficiency (upper left corner of graph).  However, these receivers encounter brick-wall limits on their achievable photon efficiencies of $\log_2 e \approx 1.44$~bits/photon (1~nat/photon) and $2\log_2 e \approx 2.89$~bits/photon (2~nats/photon), respectively. 	

Applications that require high photon efficiency typically use PPM or OOK modulation combined with a photon counting receiver.  There is no upper limit on photon efficiency achievable with PPM or OOK modulation and photon counting.  However, the maximum dimensional efficiency achievable by such systems operating at around 10~bits/photon is more than an order of magnitude lower than that achievable at the ultimate quantum limit. 	The tradeoff limits for PPM are strictly inferior to those for OOK even though PPM is regarded as a high-order modulation while OOK is binary.  The reason is that the dimensional efficiency of PPM is determined by the number of slots, and within each slot PPM is simply a constrained form of OOK.  At high photon efficiencies such as 10~bits/photon, this performance difference between PPM and OOK is very small.

Binary modulations such as OOK and BPSK cannot achieve dimensional efficiency greater than 1~bit/dimension, and the corresponding limit for PPM is only 1/2~bit/dimension.  However, the focus of this paper is in the high photon efficiency region.  At 10~bits/photon, BPSK and OOK modulations are only slightly sub-optimal with respect to the ultimate quantum limit if they could be teamed with quantum-optimal ultimate receivers.

The Dolinar receiver~\cite{Dolinar:QuantRec, Dolinar:PhDthesis} is a quantum-optimal receiver for distinguishing arbitrary binary coherent states with the objective of minimizing the uncoded error probability.  When teamed with OOK modulation, its tradeoff curve of dimensional and photon efficiencies is uniformly but insignificantly better than the corresponding tradeoff achieved by photon counting.  When teamed with BPSK modulation, the Dolinar receiver runs into the same brick-wall limit on photon efficiency (2.89~bits/photon) encountered by homodyning.
 
\section{Analysis of the Capacity Efficiency Tradeoffs}
\label{sec:CapAnalysis}
In this section, we derive the closed-form approximation presented in Fig.~\ref{fig:DIEvsPIElimits} for the DIE vs. PIE tradeoff curve attained by PPM modulation and photon counting, and we examine its asymptotic behavior for high PIE compared to that of the ultimate Holevo limit.
The resulting asymptotic formulas give rise to the two simple analytic approximations that were plotted in 
Fig.~\ref{fig:DIEvsPIElimits} alongside the true capacity tradeoff curves.

\subsection{Capacity of PPM and photon counting}
\label{sec:CapPPM+counting}
For a realizable system using $M$-ary PPM and photon counting, the capacity $C$ per PPM symbol is very simple:
\beq C = (1 - e^{-\E}) \log_2 M \eeq
where $\E$ is the average energy (number of photons) used in the one slot that contains the PPM pulse.  In this case, the photon efficiency
\beq \pie = \frac{C}{E} = \frac{1 - e^{-\E}}{\E} \log_2M \eeql{PIE(E,M):PPM}
increases without bound for large $M$ and fixed $\E$.  But the dimensional efficiency is normalized by the number of dimensions (slots) $M$,
\beq \die = \frac{C}{M} =  (1 - e^{-\E}) \frac{\log_2 M}{M} \eeql{DIE(E,M):PPM}
and this goes to zero for large~$M$.
The tradeoff between these two capacity efficiencies is typically studied by evaluating these expressions parametrically as a function of $\E$ for various fixed values of $M$.  A different $\die$ versus $\pie$ tradeoff curve is obtained for each $M$, and the optimum tradeoff is the outer envelope of all such curves.

Better insight into the behavior of the optimum tradeoff is obtained by solving for the optimum $M = M^*$ as if real values of $M^*$ were allowed.  This method yields an upper bound that is also an extremely good approximation.  To do this, we use~(\ref{PIE(E,M):PPM}) and~(\ref{DIE(E,M):PPM}) to compute 
\begin{IEEEeqnarray}{rCl}
 \log_2 \die + \pie &=& \log_2(\E\pie) - \log_2 M + \pie \nonumber \\
                                &=& \log_2(\E\pie) + \pie\left( 1 -   \frac{\E}{1-e^{-\E}}    \right)
\label{logDIE+PIE:PPM} 
\end{IEEEeqnarray}
Maximizing $\die$ for a given $\pie$ is equivalent to maximizing $ \log_2 \die + \pie $ for given $\pie$, but the latter expression is more convenient for optimization, because only $\E$ remains as a free parameter, with $M$ having been eliminated.  Differentiating the right side of~(\ref{logDIE+PIE:PPM}) with respect to $\E$ and setting the derivative to zero at $\E = \E^*$ yields an equation that can be solved explicitly for $\pie$ in terms of the optimizing value of $\E^*$:
\beq \pie =  \frac {  \left(1-e^{-{\E}^*} \right)^2 }   {  \left({\E}^*\ln 2\right) \left[1 - (1+\E^*) e^{-{\E}^*} \right]  } 
\eeql{PIEvsE:PPM}
The corresponding optimum $M=M^*$ (allowed to be real-valued) is given by:
\beq \log_2 M^* = \frac{ \E^* }{ 1-e^{-{\E}^*} }\, \pie =  \frac{1-e^{-{\E}^*} } { 1-(1+\E^*) e^{-{\E}^*} }  \frac{1}{\ln 2} 
\eeql{optMvsE:PPM}
and the resulting dimensional efficiency is:
\beq \die = \frac{ { \left(1-e^{-{\E}^*}\right)^2 \exp\left( - \frac{1 - e^{-{\E}^*} } {1 - (1+\E^*)e^{-{\E}^*} } \right) }  } 
                      {  (\ln 2) \left[1 - (1+\E^*) e^{-{\E}^*} \right]   } 
\eeql{DIEvsE:PPM}
This expression~(\ref{DIEvsE:PPM}) for the dimensional efficiency $\die$ and the previous expression~(\ref{PIEvsE:PPM}) for the photon efficiency $\pie$ together specify the optimum $\die$ versus $\pie$ tradeoff curve parametrically in terms of $\E^*$.

\subsection{Asymptotic capacity of PPM and photon counting}
In the parametric expressions~(\ref{PIEvsE:PPM}) and~(\ref{DIEvsE:PPM}) for $\pie$ and $\die$, large values of PIE correspond to small values of $\E^*$.
As $\E^* \rightarrow 0$, we have
\beq \E^* \pie \rightarrow \frac{2}{\ln 2} \eeq
and
\beq  \E^* \log_2 M^* \rightarrow \frac{2}{\ln 2} \eeq
From these two expressions we see that the optimum PPM order~$M^*$ increases exponentially with increasing $\pie$:
\beq M^* \rightarrow \tilde{M}^* \defn 2^{\pie} \eeq
Finally we see from~(\ref{logDIE+PIE:PPM}) that
\begin{IEEEeqnarray}{rCl}
 \log_2\die + \pie &\rightarrow& \log_2\left(\frac{2}{\ln 2}\right) - \frac{1}{\ln 2} \nonumber \\
                                 &=& \log_2 \left(\frac{2}{e\,\ln 2}\right) \approx 0.086
\end{IEEEeqnarray}                                
Therefore, our analytic approximation to the optimum $\die$ versus $\pie$ tradeoff curve behaves asymptotically as:
\beq \die \rightarrow\dieapproxa \defn \left(\frac{2}{e\,\ln 2}\right) \, 2^{-\pie} \approx 1.061 \times 2^{-\pie} \eeq
Thus, with PPM and photon counting, the dimensional information efficiency must fall off exponentially with increasing photon information efficiency.

\subsection{Asymptotic ultimate Holevo capacity}
At the ultimate Holevo limit, the dimensional and photon efficiencies are given by:
\begin{IEEEeqnarray}{rCl}
  \die &=& (\E+1) \log_2(\E+1) - \E \log_2 \E \nonumber \\
  \pie &=& \die/\E  
\end{IEEEeqnarray}                                
where $\E$ is the average number of photons per dimension.
For this case we compute:
\beq \log_2(\die/\pie) + \pie  =\frac{(\E+1)\log_2(\E+1)}{\E} \eeq
As $\E\rightarrow 0$,
\beq \log_2(\die/\pie) + \pie   \rightarrow\log_2 e \eeq
 So asymptotically for large $\pie$, the DIE vs. PIE tradeoff curve at the ultimate Holevo limit is given by:
\beq\die \rightarrow \dieapproxb \defn e\,\pie \, 2^{-\pie} \eeq
Even at the  ultimate limit, the dimensional efficiency $\die$ must fall off exponentially with increasing photon efficiency $\pie$, except for a multiplicative factor proportional to $\pie$.

\subsection{Usefulness of the asymptotic approximations}
The asymptotic expressions $\dieapproxa$ and $\dieapproxb$ are numerically accurate relative to the dimensional information efficiencies from which they were derived, for high enough photon information efficiency. 
The approximation~$\dieapproxa$ {\em overestimates} the dimensional efficiency achieved with ``PPM + photon counting''  by less than 10\% for photon efficiencies above 4~bits/photon.  The approximation~$\dieapproxb$ {\em underestimates} the dimensional efficiency achieved at the ultimate Holevo bound by less than 10\% for photon efficiencies above 4~bits/photon.  

The asymptotic expressions $\dieapproxa$ and $\dieapproxb$ also serve as decently accurate approximations to the dimensional information efficiencies for several of the other cases considered in Fig.~\ref{fig:DIEvsPIElimits}.  The approximation~$\dieapproxa$ {\em underestimates} the dimensional efficiency achieved with ``OOK + photon counting'' or ``OOK + Dolinar receiver'' by less than 10\% for photon efficiencies above 10~bits/photon.  The approximation~$\dieapproxb$ {\em overestimates} the dimensional efficiency achievable at the Holevo limit for the two cases of constrained modulation.  For the ``BPSK + ultimate receiver'' curve, this overestimate is within 10\% of the true curve for photon efficiencies above 7~bits/photon.  For the ``OOK + ultimate receiver'' curve, it is within 10\% of the true curve for photon efficiencies above 16~bits/photon.

Comparing the asymptotic approximations for the ultimate Holevo capacity and the capacity of PPM with photon counting, we obtain:
\beq \frac{\die(\text{ultimate Holevo})}{\die(\text{PPM + counting})} \rightarrow \frac{\dieapproxb}{\dieapproxa} 
 = \left( \frac {e^2 \ln 2} {2} \right) \pie \approx 2.561 \, \pie \eeq
This expression gives a good approximation, for high photon efficiency, to the best possible factor by which the dimensional efficiency can be improved by replacing a conventional system employing PPM and photon counting with one that reaches the ultimate Holevo limit.  This improvement factor is only linear in the photon efficiency.

\section{Capacity Efficiency of the Dolinar Receiver}
\label{CapCohDR}

In~\cite{ErkmenMoisionBirnbaum:IPNPR2009} there was a perplexing result that the Dolinar receiver, despite being precisely optimal for the binary coherent state detection problem, yielded a DIE vs. PIE tradeoff that was markedly inferior to the optimal tradeoffs obtained with photon counting of either PPM or OOK. In particular, the photon information efficiency of the Dolinar receiver appears to hit the same brick wall limit of 2.89~bits/photon as that of a homodyne receiver.  However, that conclusion results from imposing a constraint of equal {\em a~priori} probabilities on the operation of the Dolinar receiver.  In this section, we show that, when that constraint is lifted, the capacity attained by the Dolinar receiver with OOK is actually better than that achieved with photon counting, but the improvement is tiny.

\subsection{Capacity for arbitrary binary coherent states}
When a canonical Dolinar receiver is used to optimally distinguish between two equally likely coherent states with minimum probability of error, the {\em a~posteriori} probability that the receiver's observations favor the true state evolves monotonically upward with observation time as additional received energy is accumulated.  When the same receiver structure is used to optimally distinguish between two coherent states that are not equally likely {\em a~priori}, the Dolinar receiver for this case operates exactly as the canonical Dolinar receiver for equally likely states, except that it starts from the point at which the canonical receiver's {\em a~posteriori} probability has evolved to the point where it equals the {\em a~priori} probability of the more likely state.

\begin{figure}[htbp]
	\centering
	\includegraphics[width=3.5in]{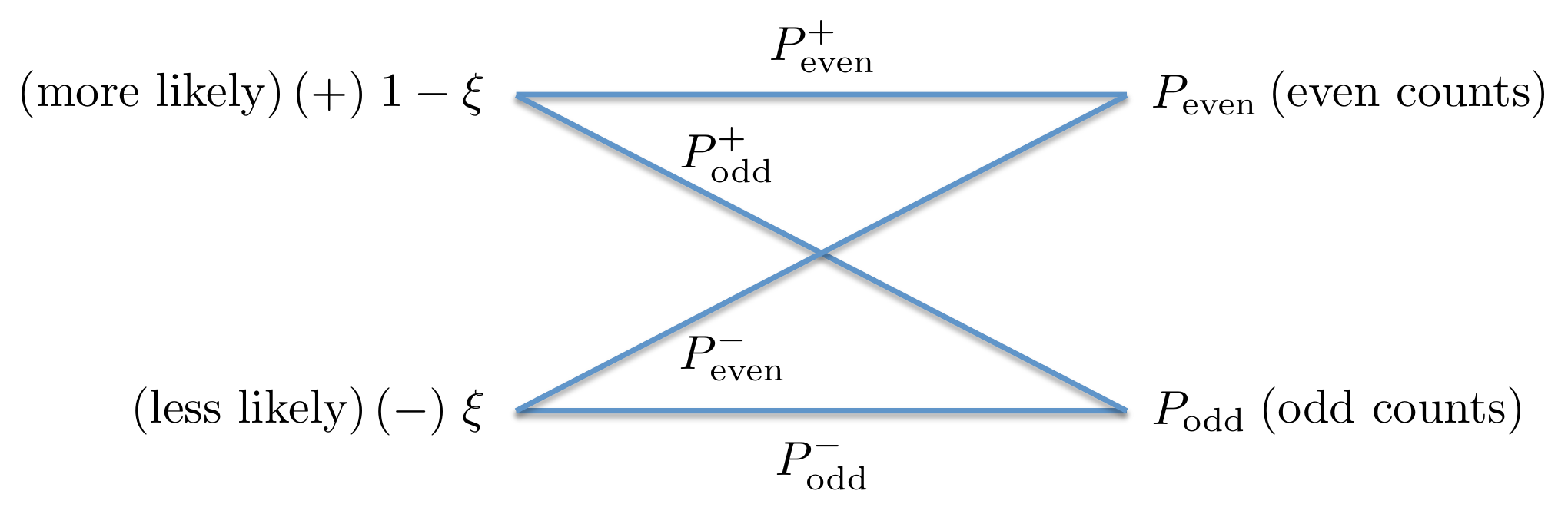}
	\caption{Binary asymmetric channel obtained by applying a Dolinar receiver measurement to binary coherent states with arbitrary {\em a~priori} probabilities.}
	\label{fig:BAC-Dolinar}
\end{figure}

When a Dolinar receiver measurement is used to distinguish two states with equal {\em a~priori} probabilities, the associated channel is a binary symmetric channel (BSC) with crossover probability equal to the receiver's error probability for the corresponding binary detection problem.  This is the case for which the formulas and curves presented in~\cite{ErkmenMoisionBirnbaum:IPNPR2009} are applicable.  When a Dolinar receiver measurement is used to distinguish two states with unequal {\em a~priori} probabilities, the associated channel may still be considered binary, but it is no longer symmetric.  In this case, the possible channel outputs divide into two categories,\footnote{No additional mutual information is obtained by resolving the possible channel outputs into more than these two categories; see Section~\ref{DR:senseofoptim}.}  as shown in Figure~\ref{fig:BAC-Dolinar}, corresponding to even and odd numbers of observed photon counts, and the conditional probabilities of observing even or odd counts are different for the two channel inputs.

The conditional probabilities of obtaining even counts were derived in eqs.~(6.34) and~(6.35) of~\cite{Dolinar:PhDthesis}, but here we will express them more conveniently.  
Let $\ket{\psi_0}$ and $\ket{\psi_1}$ denote the two coherent states, and let $s \defn \magsqr{\braket{\psi_0}{\psi_1}}$ denote their
overlap (squared inner product).  If $\xi$ denotes the {\em a~priori} probability of the less likely state, then the
conditional probabilities of getting even and odd numbers of counts can be written as:
\begin{IEEEeqnarray}{rCl}
\Pev^+   &=&     \frac{1}{2} \left( 1 + \frac {1-2\xi s}     {\sqrt{1 - 4\xi(1-\xi)s}}    \right)    \nonumber \\
\Podd^+ &=&    \frac{1}{2} \left( 1   - \frac {1-2\xi s}     {\sqrt{1 - 4\xi(1-\xi)s}}    \right)   \nonumber \\
\Pev^-    &=&     \frac{1}{2} \left( 1 - \frac {1-2(1-\xi) s} {\sqrt{1 - 4\xi(1-\xi)s}}    \right)   \nonumber \\
\Podd^-  &=&    \frac{1}{2} \left(1 + \frac {1-2(1-\xi) s} {\sqrt{1 - 4\xi(1-\xi)s}}   \right)
\label{DolBACprobabilities}
\end{IEEEeqnarray}  
where the conditioning events corresponding to the more probable and less probable states are denoted by the superscripts~$+$ and~$-$, respectively.
The probability of error achieved by this receiver is obtained by averaging the two crossover probabilities in Fig.~\ref{fig:BAC-Dolinar}:
\beq \Pe =   \xi \Pev^-  +   (1-\xi) \Podd^+   \eeq
This error probability achieves the quantum limit for distinguishing between two states, called the Helstrom bound~\cite{helstrom-bound}:
\begin{equation}
	\Pe = \frac{1}{2}\left( 1-\sqrt{1 - 4 \xi (1-\xi) s} \, \right).
\end{equation}
                              
In the special case when $\xi = 1/2$, these four conditional probabilities characterize a BSC with crossover probability 
\beq  \Podd^+=   \Pev^- =   \frac{1}{2} \left( 1 - \sqrt{1 - s} \, \right) \eeq
In the more general case, the mutual information $I(X;Y)$ conveyed across the asymmetric channel shown in Fig.~\ref{fig:BAC-Dolinar} is:
\beq I(X;Y) = H_2(\Podd) - \left[ \xi H_2(\Pev^-) + (1-\xi) H_2(\Podd^+) \right] \eeql{mutualinfo-Dolinar}
where $ H_2(x) = -x\log_2 x - (1-x)\log_2(1-x) $ is the binary entropy function, and $\Podd$ is the unconditional probability of observing odd counts:
\beq \Podd = \xi \Podd^- + (1-\xi) \Podd^+ \eeq
The mutual information expression~(\ref{mutualinfo-Dolinar}) is valid for the channel obtained by applying the Dolinar receiver measurement to any pair of coherent states.  However, this channel is peculiar, because its crossover probabilities are dependent on the {\em a~priori} probabilities of its inputs.  
The maximum mutual information {\em per channel use} is attained for $\xi = 1/2$, and this yields the capacity {\em per channel use} of a BSC as stated in~\cite{ErkmenMoisionBirnbaum:IPNPR2009}:
\beq C = 1 - H_2(\Pe)  \qquad \text{(bits per channel use)} \eeq
For BPSK-modulated coherent states, the maximum mutual information {\em per photon} is also attained for $\xi = 1/2$, because in this case the two states have equal energy. 

\subsection{Capacity for OOK modulation}
 If the two coherent states are not equally energetic,
the maximum mutual information {\em per photon} is generally attained for $\xi < 1/2$, where $\xi$ is the {\em a~priori} probability of the more energetic state.  
OOK modulation produces the most disparate state energies.
In this case, the overlap between the ``ON'' and ``OFF'' states is $s = e^{-\E/\xi}$, where $\E/\xi$ is the average number of photons in the ``ON'' pulse.  The optimum channel input probability
$\xi$ for maximizing mutual information {\em per photon} approaches~0 in the limit of large photon information efficiency.  Closed-form expressions are not available.  The DIE vs. PIE tradeoff curve for this case in Fig.~\ref{fig:DIEvsPIElimits} was obtained by maximizing~(\ref{mutualinfo-Dolinar}) numerically over $\xi$ for given $\E$ to obtain $C(\E)$, then varying~$\E$ parametrically to trace out $\die = C(E)$ vs. $\pie = C(E)/E$. This curve is uniformly better than the corresponding DIE vs. PIE tradeoff curve obtained for OOK modulation with photon counting, but only by a minuscule amount.

\section{Sense of Optimality of the Dolinar Receiver}
The Dolinar receiver is known to be the optimal hard-decision measurement on an arbitrary binary coherent-state alphabet. 
In this section, we show that it is also an optimal soft-decision receiver, at least for BPSK, in the sense of {\em maximizing the mutual information} obtained from the measurement, within a class of causal optical feedback receivers.  For BPSK, we also show that the Dolinar receiver achieves the capacity of an ultimate receiver for a certain {\em reciprocal channel} with a complementary signaling geometry.

\subsection{Optimality with respect to mutual information}
\label{DR:senseofoptim}
\begin{figure}[!t]
\centering
\includegraphics[width=3in]{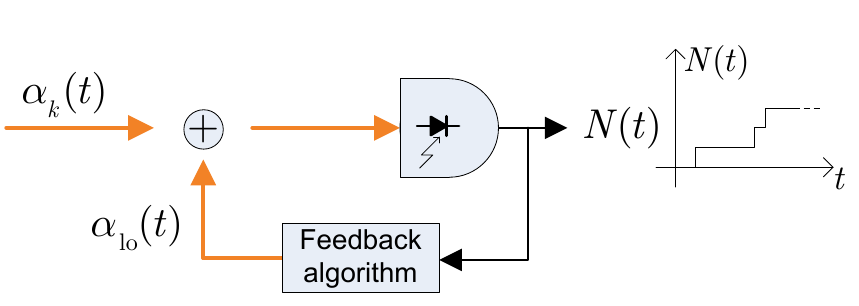}
\caption{The general structure of an optical feedback receiver, which modifies a local oscillator field $\alpha_{\lo}(t)$ as a function of the photodetector output $N(t)$. The orange thicker lines indicate optical fields and the thinner black lines represent electrical signals. }
\label{FBRec}
\end{figure}

Consider the optical detection system in Fig.~\ref{FBRec}, in which the received optical field is first displaced by an ideal coherent-state local oscillator, and subsequently observed with an ideal (unity quantum efficiency, no dark current) photon-counting photodetector. 
We assume that the incoming signal is constant over $t \in (0,T]$, and its value is randomly selected from an alphabet of $\mathcal{K}$ coherent states,  $\{\ket{\alpha_{k}}, k = 1, 2, \dots, \mathcal{K}\}$, with {\em a~priori} probabilities $\{p_k, k = 1, 2, \dots, \mathcal{K}\}$. Therefore, the output of the photodetector is a conditionally Poisson counting process $N(t)$, with rate $\lambda(t) = | \alpha_{k} + \alpha_{\lo}(t)|^{2}$ for $t\in (0,T]$. We ignore bandwidth and dynamic range limitations in the feedback path and assume that the local oscillator field (amplitude and phase) can be varied instantaneously, based on the counting process output from the photodetector. The objective function to be maximized by the local oscillator is the mutual information between the random variable $k$ and the counting process $N(t)$ over the time window $[0,T]$. In particular, the objective is to maximize 
\begin{equation}
I(K; \{N(t): 0<t\leq T\})\,, \label{MI:Gl}
\end{equation}
by choosing the local oscillator field $\alpha_{\lo}(t)$ using the only information that the receiver has, i.e., the photon count times observed in $N(t)$.\footnote{Note that the local oscillator field must depend on the counting process causally.} We assume that $N(0)=0$ with probability~$1$, with no loss of generality.

We will first prove that maximizing the global mutual information given in \eqref{MI:Gl} is equivalent to incrementally maximizing the mutual information in the time window $(t,t+\Delta t]$, as $\Delta t \rightarrow 0$, conditioned on the observations up to and including time $t$. Let us begin by expressing \eqref{MI:Gl} as the following limit,
\begin{equation}
I(K; \{N(t): 0<t\leq T\}) = \lim_{\dT\rightarrow 0} I(K; \{N_{m'}\}_{0}^{\lfloor T/\dT \rfloor})\,, \label{MI:lim}
\end{equation}
where $N_{m} = N((m+1) \dT) - N(m \dT)$, and $\{N_{m'}\}_{j}^{m} \equiv \{N_j,N_{j+1},\dots,N_{m}\}$ for $j \leq m$. We can now use the chain rule for mutual information to expand the argument of the limit as
\begin{equation}
I(K; \{N_{m'}\}_{0}^{\lfloor T/\dT \rfloor}) = \sum_{m=0}^{\lfloor T/\Delta T \rfloor} \dT \frac{I(K;N_{m} | \{ N_{m'}\}_{0}^{m-1})}{\dT}\,. \label{MI:exp}
\end{equation}
The statistical regularity of a Poisson process ensures that the second term inside the summation converges as $\dT$ converges to $0$. Hence, substituting \eqref{MI:exp} into \eqref{MI:lim}, and recognizing that the local oscillator field $\alpha_{\lo}(t)$ is a (causal) function of the counting process realization, we obtain
\begin{multline}
\max_{\alpha_{\lo}(t)} I(K; {N(t): 0<t\leq T}) = \int_{0}^{T} {\rm d}t  \, {\rm E}_{N} \left [\max_{\alpha_{\lo}(t)} \right. \\ \left. \lim_{\dT\rightarrow 0} \frac{i(K; N(t+\dT)-N(t)|\{n(\tau):\tau \in (0,t]\})}{\dT}  \right ] \,, \label{MI:Incr}
\end{multline}
where $n(t)$ denotes a particular realization of the photodetector output $N(t)$, and $i(X;Y|z)$ is the mutual information between $X$ and $Y$, {conditioned} on a \emph{particular realization} of the random variable $Z$.\footnote{Note that the conditional mutual information is given by $I(X;Y|Z) = {\rm E}_{Z}[i(X;Y|z)]$~\cite{CoverThomas:InfThy}.} Equation \eqref{MI:Incr} shows that the optimal local oscillator that maximizes the mutual information between the input symbol $k$ and the photodetector output over $0<t\leq T$ can be chosen at each time instant $t$, such that it incrementally maximizes the conditional mutual information at the next time instant, given the observations up to and including $t$.

Next, we turn our attention to the numerator of the limit in \eqref{MI:Incr}, given by
\begin{multline}
i(K; N(t+\dT)-N(t)|\{n(\tau); \tau \in (0,t]\}) = \\
H(N(t+\dT) - N(t) | \{n(\tau); \tau \in (0,t]\}) \\
 - H(N(t+\dT) - N(t) | K, \{n(\tau); \tau \in (0,t]\})\,,
\end{multline}
where $H(\cdot)$ is the well-known (discrete) entropy function~\cite{CoverThomas:InfThy}. Because $N(t)$ is a Poisson process when conditioned on $K$ (and $\alpha_{\lo}(t)$), $N(t+\dT) - N(t)$ is a conditionally-Poisson random variable with mean approximately $\dT |\alpha_{\lo}(t)+\alpha_{k}|^{2}$, and a compound Poisson random variable when conditioned on $\alpha_{\lo}(t)$ alone. Using tight bounds on the entropy of Poisson and compound Poisson random variables~\cite{Adell:PoissonEntropy}, it can be shown that
\begin{multline}
\lim_{\dT \rightarrow 0} \frac{i(K; N(t+\dT)-N(t)|\{n(\tau); \tau \in (0,t]\})}{\dT} \\
= -\overline{\lambda} \log\overline{\lambda} + \sum_{k=1}^{K} p_{k} \lambda_{k} \log\lambda_{k}\,, \label{MI:cond}
\end{multline}
where $\lambda_{k}\equiv |\alpha_{\lo}(t)+\alpha_{k}|^{2}$ for $k \in \{1,2,\dots,\mathcal{K}\}$, $\overline{\lambda} \equiv \sum_{k} p_{k} \lambda_{k}$, and we have suppressed the time-dependence of the variables to avoid clutter. It is worthwhile to point out that the right-hand side of \eqref{MI:cond} is a convex function of the $\lambda_{k}$'s and therefore is non-negative, as required. The right hand side of Eq.~\eqref{MI:cond} is in a form that can be maximized in terms of $\alpha_{\lo}$. In general, $\alpha_{\lo}\equiv a \exp(i\phi)$ is complex, thus the maximization must be carried out over both the amplitude and the phase. The optimal solution must therefore satisfy,
\begin{align}
\sum_{k} p_k \log\left(\frac{\lambda_{k}}{\overline{\lambda}} \right ) (a + |\alpha_{k}| \cos(\phi-\phi_{k})) &= 0\\
\sum_{k} p_k \log\left(\frac{\lambda_{k}}{\overline{\lambda}} \right ) \sin(\phi-\phi_{k}) &= 0\,,
\end{align}
where $\phi_{k} \equiv \angle\alpha_{k}$. The solution to these equations for general $\{ \alpha_{k} \}$ is nontrivial. It can be noted, however, that if all $\alpha_{k}$ have common phase (up to a $\pi$ phase shift), then the local oscillator is also either in phase or out-of-phase with the constellation. 

In particular, in a BPSK scenario with a constellation $\{\ket{-\alpha}, \ket{\alpha}\}$, the solution to these equations can be found analytically, and perhaps unsurprisingly, the optimal local oscillator turns out to be \emph{identical} to the local oscillator that minimizes the probability of error in classifying the signal state, i.e., the Dolinar receiver~\cite{Dolinar:QuantRec}. This result shows that, for BPSK at least, the Dolinar receiver is the optimal soft-decision receiver within the class of causal optical feedback receivers. But this does not imply that higher mutual information cannot be achieved with a receiver outside of this class.

\subsection{Reciprocal channel optimality}
For BPSK modulation, the capacity (bits) achieved by the Dolinar receiver is given in terms
of the energy $E$ (photons) of each BPSK-modulated coherent state $\ket{\psi_0}$, $\ket{\psi_1}$:
\beq C_{\rm Dol+BPSK}  = 1- H_2\left(\frac{1}{2}\left(1-\sqrt{1-e^{-4E}} \right)\right) \eeq
where $H_2(\cdot)$ is the binary entropy function.
The Holevo capacity for BPSK-modulated coherent states is:
\beq C_{\rm Hol+BPSK}  = H_2\left(\frac{1}{2}\left(1-e^{-2E}\right)\right) \eeq
Let $\overlap$ denote the overlap between the two BPSK-modulated coherent states.
\beq \overlap = \magsqr{\braket{\psi_0}{\psi_1}} = e^{-4E}  \qquad \text{(for BPSK)} \eeql{BPSKoverlap}
Writing the two capacity expressions in terms of $\overlap$, we see that:
\beq C_{\rm Dol+BPSK}(\overlap)  = 1- H_2\left(\frac{1}{2}(1-\sqrt{1-\overlap})\right)\eeq
and
\beq C_{\rm Hol+BPSK}(\overlap)  = H_2\left(\frac{1}{2}(1-\sqrt{\overlap})\right) \eeq
Thus, we have the result that:
\beq C_{\rm Hol+BPSK}(\overlap) + C_{\rm Dol+BPSK}(1-\overlap) = 1~{\rm bit} \eeq
In other words, the ultimate receiver and the Dolinar receiver for BPSK are ``complementary'' or ``reciprocal'' in that the total potential capacity of any binary modulation (1~bit) is obtainable as the sum of the capacities of the two receivers given complementary signaling geometries, one with $\magsqr{\braket{\psi_0}{\psi_1}} = \overlap$ and one with $\magsqr{\braket{\psi_0}{\psi_1}} = 1-\overlap $. This intriguing property is exact, not an approximation.  It is similar in concept to a reciprocal channel relation between variable and check node operations in decoders for low-density parity-check (LDPC) codes,  that is used to accurately approximate density evolution~\cite{Chung:PhDthesis}.  But its significance in the current context is unclear.

\section{Dolinar receiver with adaptive priors}
\label{sec:adaptive-Dolinar}
The previous two sections focused on computing the capacity of the Dolinar receiver applied to a single symbol represented by one of two coherent states.
In this section, we extend the Dolinar receiver to make {\em adaptive measurements} on a {\em coded sequence} of coherent state symbols. Information from previous measurements is used to adjust the {\em a~priori} probabilities of the next symbols. 

\subsection{Definition of the adaptive Dolinar receiver}
Consider a communication channel with binary modulation.  The transmitter puts each quantum mode into either state $\ket{\psi_0}$ or state $\ket{\psi_1}$.  The overlap of the two states is $\magsqr{\braket{\psi_0}{\psi_1}}\equiv s$.  When $s\neq0$, the two states are not orthogonal, and therefore are not perfectly distinguishable.  We will consider a channel which has no effect on the transmitted states (the identity map channel), so that the receiver may perform measurements directly on the transmitted states.  

Consider a receiver that yields a random measurement result $Y\in\{0,1\}$.  The receiver may be defined by two Hermitian operators $\{\qop{M_0}, \qop{M_1}\}$, where the probability of measurement result $y$ given state $\ket{\psi_x}$ is given by 
\begin{equation}
	P(Y=y|X=x)=\bra{\psi_x}\qop{M}_y\ket{\psi_x}.  
\end{equation}
Since the two possible outcomes are collectively exhaustive, $\qop{M}_0+ \qop{M}_1 = \qop{I}$, where $\qop{I}$ is the identity operator on the space spanned by $\{\ket{\psi_0},\ket{\psi_1}\}$.

The Dolinar receiver~\cite{Dolinar:QuantRec, Dolinar:PhDthesis, dolinar-receiver-experiment} implements a projective measurement ($\qop{M}_y \qop{M}_{y'} = \delta_{y,y'}\qop{M}_y$) on a single optical mode.  
The measurement operators $\qop{M}_y = \qop{M}_y(\xi)$ are functions of the \emph{a~priori} probability $\xi$ that state $\ket{\psi_1}$ is transmitted.
For the Dolinar receiver measurement, the probabilities of the two possible outcomes $y$, given the two possible states $\ket{\psi_x}$, are the four conditional probabilities evaluated in~(\ref{DolBACprobabilities}) and depicted in Fig.~\ref{fig:BAC-Dolinar}:
\begin{IEEEeqnarray}{rCl}
P(Y=0|X=0;\xi) &=& \Pev^+     \nonumber \\
P(Y=1|X=0;\xi) &=& \Podd^+   \nonumber \\
P(Y=0|X=1;\xi) &=& \Pev^-     \nonumber \\
P(Y=1|X=1;\xi) &=& \Podd^- 
\label{DolBACprobabilities:XY}
\end{IEEEeqnarray}  

The Dolinar receiver measures a single mode of the optical field.  We now consider $n$ optical modes, which may differ in their spatial, temporal, or polarization dimensions.  The transmitter prepares mode $j$ (where $j\in\{1,2,\dots n\}$) in state $\ket{\psi_{x_j}}$, where $x_j\in\{0,1\}$ for binary modulation.  The quantum state of the system may then be represented as the product state
\begin{equation}
	\ket{\Psi_{\vec{x}}} = \bigotimes_{j=1}^{n}\ket{\psi_{x_j}}
\end{equation}
We  represent the $n$ binary digits $\{x_j\}$ as the vector $\vec{x}$; there are $2^n$ possible such vectors, which we will denote as $\{\vec{x}^{(0)}, \vec{x}^{(1)}, \dots, \vec{x}^{(2^n)}\}$.  Again we assume that the channel is an identity map, so the receiver may perform measurements directly on the transmitted states.  

We now describe a new receiver for the $n$-mode channel.  This receiver is an extension of the Dolinar receiver which makes use of information received on some modes to adapt the measurement of other modes.  The new receiver performs measurements of each mode sequentially and consecutively (the measurement of one mode is completed before the measurement of the next mode begins).  This could be further extended to include measurements which occur in parallel on multiple modes, so that partial measurement results could be used in an adaptive algorithm; however, we do not explore this possibility here.  The new receiver performs a Dolinar measurement on each mode, but with a variable parameter $\xi_j$ that depends on the outcome of previous measurements.  The $2^n$ measurement operators are thus of the form
\begin{equation}
	\qop{\mathcal{M}}^{(l)} = \bigotimes_{j=1}^{n} \qop{M}_{y_{j}^{(l)}}(\xi_j), \quad l\in \{0,1,\dots 2^n\}
\end{equation}
where $\{\vec{y}^{(0)}, \vec{y}^{(1)},\dots \vec{y}^{(2^n)}\}$ is the set of possible measurement outcome sequences.  These outcomes are completely exhaustive, so that 
\begin{equation}
	\sum_{l=1}^{2^n} \qop{\mathcal{M}}^{(l)} =  \qop{\mathcal{I}}
\end{equation}
where $\qop{\mathcal{I}}$ is the identity operator on the space spanned by the $2^n$ state vectors $\{ \ket{\Psi_{\vec{x}^{(0)}}}, \ket{\Psi_{\vec{x}^{(1)}}}, \dots \ket{\Psi_{\vec{x}^{(2^n)}}} \}$.  We take $\xi_j$ to be a function of previous measurement results, so $\xi_j =\xi_j(y_0,y_1,\dots y_{j-1})$.

The measurement result of mode $j$ is given by $Y_j\in\{0,1\}$, and $\vec{Y}$ represents the $n$ binary digits of $\{Y_j\}$.  The conditional probability of a sequence of measurement outcomes may be written as 
\begin{equation}
	\label{eq:conditional-total}
	P(\vec{Y}=\vec{y}|\vec{X}=\vec{x}) = \prod_{j=1}^{n} P\left( Y_j=y_j |X_j=x_j ; \xi_j \right)
\end{equation}
where the conditional probability of each outcome is given by 
\begin{equation}
	\label{eq:conditional-part}
	P\left( Y_j=y_j |X_j=x_j ; \xi_j \right) = \bra{\psi_{x_j}}\qop{M}_{y_j}(\xi_j)\ket{\psi_{x_j}}.
\end{equation}
and $\qop{M}_{y}(\xi)$ is the single-mode Dolinar operator for outcome $y$ with parameter $\xi$.

The conditional probabilities of $\vec{Y}$ given $\vec{X}$ are completely described by equations~(\ref{DolBACprobabilities:XY}), (\ref{eq:conditional-total}), and (\ref{eq:conditional-part}), once the function $\xi_j(y_0,y_1,\dots y_{j-1})$ is specified.  We select the function
\begin{equation}
	\label{eq:xi-rule}
	\xi_j = \sum_{l=1}^{2^n}x_j^{(l)} g_j^{(l)}.
\end{equation}
where the values of $g_j^{(l)}$ obey $0 \leq g_j^{(l)} \leq 1$ and are defined by the iterative relation
\begin{equation}
	\label{eq:g-rule}
	g_j^{(l)} = g_{j-1}^{(l)} P(Y_{j-1}=y_{j-1} | X_{j-1} = x_{j-1}^{(l)};\xi_{j-1})/S_j
\end{equation}
where $S_j$ is a normalization constant chosen such that 
\begin{equation}
	\sum_{l=1}^{2^n} g_j^{(l)}=1, \quad \forall j.
\end{equation}
The initial values are given by the \emph{a~priori} probabilities of sending each string,
\begin{equation}
	\label{eq:g-init}
	g_0^{(l)} = P(\vec{X}=\vec{x}^{(l)}).
\end{equation}
From the receiver's perspective, $g_j^{(l)}$ represents the probability that string $\vec{x}^{(l)}$ is the transmitted string, conditioned on the measurement results $\{y_0, y_1,\dots y_{j-1}\}$.


If all messages $\vec{x}^{(l)}$ have equal \emph{a~priori} probabilities, then equations~(\ref{eq:xi-rule}), (\ref{eq:g-rule}), and~(\ref{eq:g-init}) lead to the trivial feedback rule $\xi_j = 1/2, \forall j$.  In order to investigate the performance of this receiver, we imposed a code constraint on the allowed transmitted sequences.  That is, we selected
\begin{equation}
	P(\vec{X}=\vec{x}^{(l)}) = 
	\begin{cases}
		1/2^k & \vec{x}^{(l)} \in \codeset \\
		0 & \vec{x}^{(l) }\notin \codeset
	\end{cases}
\end{equation}
where $\codeset$ is a set of sequences which obey a code constraint, and the cardinality of $\codeset$ is $2^k$.  The code is described as an $(n,k)$ code, since $k$~bits of information are represented in the sequence of length $n$.

\subsection{Adaptive Dolinar receiver applied to small codes}
The first code we consider is the $(3,2)$ parity check code, where the members of the set $\codeset$ satisfy $x_3 = x_1 \oplus x_2$.  We take the case of BPSK modulation, where the physical states used are optical coherent states, $\ket{\psi_0} = \ket{\alpha}$ and $\ket{\psi_1} = \ket{-\alpha}$.  From~(\ref{BPSKoverlap}), the overlap of the two BPSK-modulated coherent states is $s=\exp(-4{\E})$, where $\E$ is the average number of photons used in each mode.

\begin{figure}[!t]
	\centering
	\includegraphics[width=2.5in]{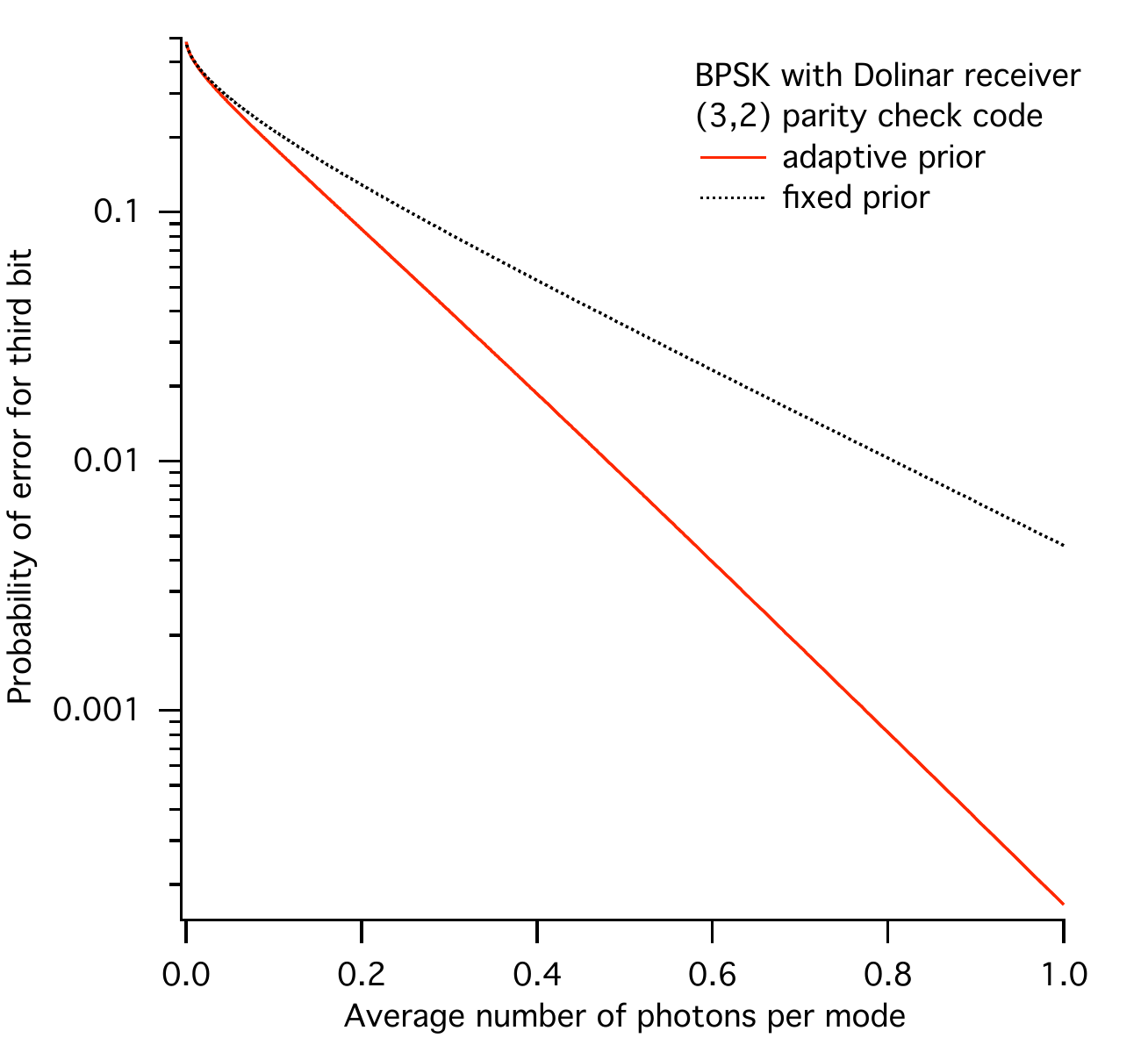}
	\caption{Probability of measurement error in the third bit $P(Y_3\neq X_3)$ vs.\ the average number of photons per mode $\Nph$ assuming BPSK modulation.  The black dotted line indicates the performance of a Dolinar receiver with fixed parameter $\xi=1/2$, while the solid red line is for the new adaptive receiver.  The message constraint is the (3,2) parity check code.}
	\label{fig:parity_error}
\end{figure}

In Figure~\ref{fig:parity_error}, we plot the probability of error in the third bit, $P(Y_3\neq X_3)$ vs.\ $\Nph$.  For comparison, we also plot the same quantity for a Dolinar receiver with fixed parameter $\xi = 1/2$.  It is apparent that the adaptive receiver has a lower probability of error, particularly for higher photon numbers. 

\begin{figure}[!t]
	\centering
	\includegraphics[width=2.5in]{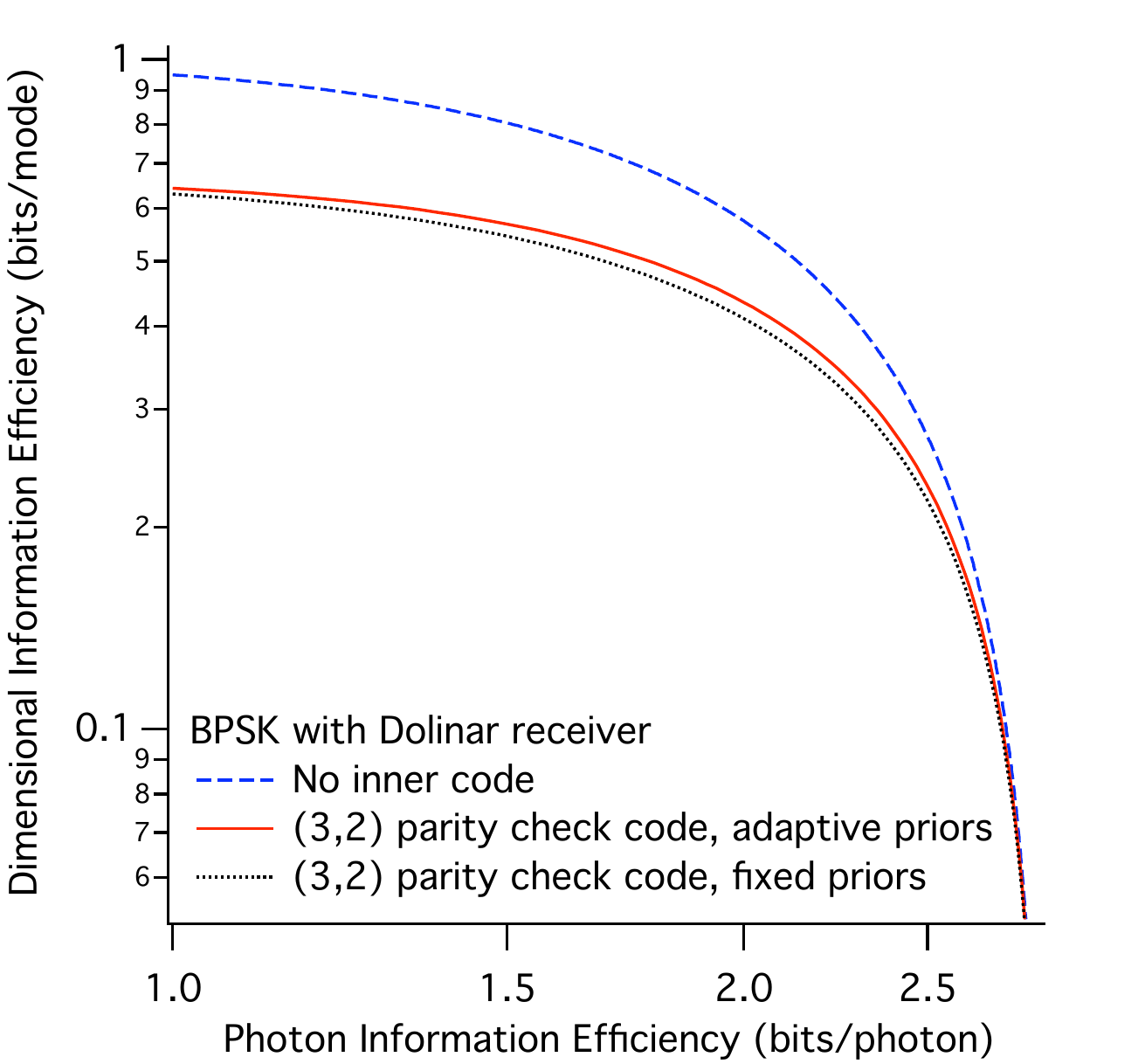}
	\caption{Dimensional information efficiency $\die$ vs.\ photon information efficiency $\pie$ for three BPSK receivers.  The dashed blue line is an uncoded system with a Dolinar receiver.  The dotted black line has the $(3,2)$ parity check code constraint and uses a Dolinar receiver with fixed parameter $\xi=1/2$.  The solid red line uses the $(3,2)$ parity check and the adaptive receiver.} 
	\label{fig:parity_pie_die}
\end{figure}

Probability of error, however, is not the only metric of interest in the communication system.  The dimensional information efficiency $\die$ and photon information efficiency $\pie$ are given by
\begin{IEEEeqnarray}{rCl}
	\label{eq:die}
	\die & = & I(\vec{X};\vec{Y})/n \nonumber \\
	\label{eq:pie}
	\pie & = & I(\vec{X};\vec{Y})/(n\Nph)
\end{IEEEeqnarray}
where $I(\vec{X};\vec{Y})$ is the mutual information between the random binary strings $\vec{X}$ and $\vec{Y}$. 

Figure~\ref{fig:parity_pie_die} shows these quantities.  For comparison, the same quantities are plotted for a transmitter with the $(3,2)$ parity check code constraint and a receiver without feedback ($\xi_j = 1/2, \forall j$), and for a transmitter without a code constraint and a receiver without feedback.  The adaptive receiver performs slightly better than the fixed receiver with the code constraint, though the gap is quite small.  The large improvements in error rate evident in Figure~\ref{fig:parity_error} have only a modest effect on $\die,\pie$ since they occur at comparatively large $\Nph$.  In this regime, the mutual information for both the fixed and adaptive receiver is approaching its maximum value of $k=2$~bits, corresponding to the asymptotic limit $\lim_{\pie \to 0} \die \to k/n = 2/3$.  On the other hand, the system without the code constraint approaches $\die = 1$~bit/mode due to the assumption of binary modulation.  So in the high $\Nph$ limit, where $\pie$ is small, the uncoded system has a higher $\die$.

In the opposite limit, as $\Nph\to 0$, the mutual information approaches zero since the states become indistinguishable, and hence $\die \to 0$.  In this regime the adaptive receiver is approaching the same measurement strategy as the fixed receiver, as seen by the converging error probabilities towards the left of Figure~\ref{fig:parity_error}.  The $\pie$ of all three receivers shown in Figure~\ref{fig:parity_pie_die} appear to approach the same limit of 2.89~bits/photon.

\begin{figure}[!t]
	\centering
	\includegraphics[width=2.5in]{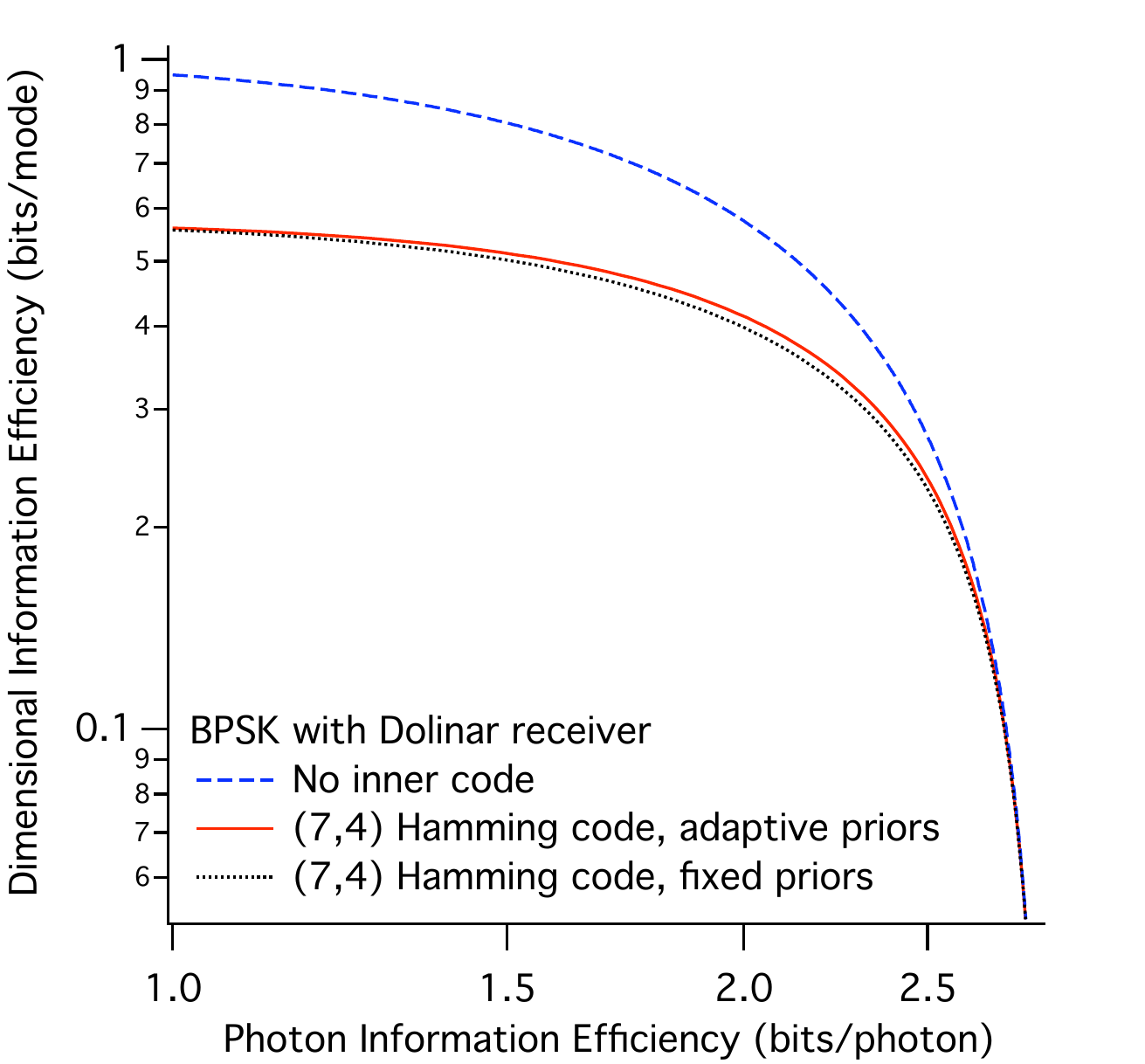}
	\caption{Dimensional information efficiency $\die$ vs.\ photon information efficiency $\pie$ for three BPSK receivers.  The dashed blue line is an uncoded system with a Dolinar receiver.  The dotted black line has the $(7,4)$ Hamming code constraint and uses a Dolinar receiver with fixed parameter $\xi=1/2$.  The solid red line uses the $(7,4)$ Hamming code and the adaptive receiver.}
	\label{fig:hamming_pie_die}
\end{figure}
In order to see the dependence on the choice of code, we also compute $\die$ and $\pie$ for the same receivers with the elements of $\codeset$ defined by the $(7,4)$ Hamming code constraints
\begin{IEEEeqnarray}{rCl}
	\label{eq:hamming}
	x_5 & = & x_1 \oplus x_2 \oplus x_3,\nonumber \\
	x_6 & = & x_1 \oplus x_2 \oplus x_4,\nonumber \\
	x_7 & = & x_1 \oplus x_3 \oplus x_4.
\end{IEEEeqnarray}
The results, shown in Figure~\ref{fig:hamming_pie_die}, are similar to those for the $(3,2)$ parity check code.  Again, the adaptive receiver slightly outperforms the fixed receiver given the code constraint, but does not exceed the efficiency of the unconstrained system.

\subsection{Non-improvement of capacity with adaptivity}
	Let us denote $C^{(1)}$ as the single symbol measurement capacity of a channel, i.e., the measurement is identical from one symbol to the next, and it is optimized (over all possible single symbol measurements) to maximize the capacity. We define the $n$-symbol measurement capacity by extension as the capacity maximized over all possible measurements that can be done jointly over $n$ symbols.
	We denote an $n$-symbol-spanning measurement as $\qop{\mathcal{M}}^{(n)}$, where the eigenspaces of the operator define a probability operator-valued measurement (POVM), and the eigenvectors are the corresponding measurement outcomes. An $n$-symbol measurement is called \emph{adaptive} if it can be represented as
	\begin{equation}
		\label{eq:adaptive}
		\qop{\mathcal{M}}_{A}^{(n)} = \bigotimes_{i=1}^{n}\qop{M}^{(1)}_{i}(y_0,\dots,y_{i-1})\,,
	\end{equation}
where $y_0,\dots,y_{i-1}$ indicates the past outcomes from the sequence of measurements. Suppose we denote the $n$-symbol \emph{adaptive} measurement capacity as $C_{A}^{(n)}$. It has been shown~\cite{Fujiwara:AdaptiveCapacity} that 
\begin{equation}
C_{A}^{(n)} = C^{(1)}\,,
\end{equation}
	i.e., the channel capacity with any single-symbol measurement strategy that adaptively changes to previous outcomes \emph{cannot} perform better than the best single-symbol nonadaptive measurement. It is also known \cite{GiovannettiGuhaEtAl:PureLossCapacity} that a single-symbol measurement is bounded away from the Holevo capacity whenever the density matrices at the output of the channel do not commute ($s\neq 0$ in our case). However, the gap between $C^{(1)}$ and the Holevo bound is not known. Furthermore, short of a brute-force search, there is in general no method to find the optimal $\qop{M}^{(1)}$. 
	Thus, there remains the possibility that an adaptive receiver can improve the capacity beyond the suboptimal measurement strategies that are currently known. Nonetheless, in light of the negative results that have been obtained in Section~\ref{sec:adaptive-Dolinar}, we conjecture that adaptive measurements of the type in (\ref{eq:adaptive}) will not significantly close the gap from known receivers to Holevo.  Instead, we suspect that measurement operators will be needed which are not factorable between modes.

\section{Conclusion}
We developed a closed-form parametric formula for the DIE vs. PIE tradeoff for a system using PPM and photon counting, and we presented analytic asymptotic expressions for this case and for the ultimate quantum limit.  We showed that any system using PPM and ideal photon counting must fall short of the maximum achievable DIE by a factor that increases linearly with PIE for high PIE.  

We worked out the mutual information resulting from a Dolinar receiver measurement for general binary coherent states with unequal {\em a~priori} probabilities, and used this result to numerically compute the DIE vs. PIE tradeoff for this receiver used with on-off keying (OOK).  But this provided only a minuscule improvement compared to OOK with photon counting.
We derived an adaptive rule for an additive local oscillator that maximizes the mutual information between a receiver and a  
transmitter that selects from a set of coherent states.  For binary phase-shift keying (BPSK) at least, this is equivalent to the operation of the Dolinar receiver.  For BPSK, we also found that the Dolinar receiver achieves the capacity of an ultimate receiver for a certain reciprocal channel with a complementary signaling geometry.

We attempted to beat the Dolinar receiver's DIE vs. PIE tradeoff for single symbols by applying an adaptive version of this receiver to multiple coded symbols, and then adjusting the {\em a~priori} probabilities of the parity symbols based on measurements of the information symbols.  This approach yields a tiny improvement over the capacity achieved by the Dolinar receiver subject to the code constraint, but the capacity worsens relative to that of the same modulation and receiver without the code constraint.

\section*{Acknowledgment}
This research was supported by the DARPA InPho program under contract number JPL 97-15402, and was carried out by the Jet Propulsion Laboratory, California Institute of Technology, under a contract with the National Aeronautics and Space Administration. 


%

%
%
%
%
%


\end{document}